\documentclass{kluwer}    

\usepackage[dvips]{graphicx}

\newdisplay{guess}{Conjecture}

\begin{document}                                                                                   
\begin{article}
\begin{opening}         
\title{The evolution of disk galaxies in clusters\thanks{
Partially based on observations collected
at the European Southern Observatory, Chile (ESO N$^{\rm o}$~66.A-0376)}
} 
\author{Alfonso  \surname{Arag{\'o}n-Salamanca}} 
\author{Bo  \surname{Milvang-Jensen}}  
\institute{School of Physics and Astronomy, University of Nottingham, UK}
\author{George \surname{Hau}}
\institute{European Southern Observatory, Chile}
\author{Inger \surname{J{\o}rgensen}}
\institute{Gemini Observatory, USA}
\author{Jens \surname{Hjorth}}
\institute{University of Copenhagen, Denmark}

\runningauthor{Arag{\'o}n-Salamanca et al.}
\runningtitle{The evolution of disk galaxies in clusters}

\date{November 30, 2001}

\begin{abstract}

We are carrying out a programme to measure the evolution of the stellar and
dynamical masses and $M/L$ ratios for a sizeable sample of
morphologically-classified disk galaxies in rich galaxy clusters at $0.2 < z <
0.9$. Using FORS2 at the VLT we are obtaining rotation curves for the cluster
spirals so that their Tully--Fisher relation can be studied as a function of
redshift and compared with that of field spirals.   We already have rotation
curves for $\sim$10 cluster spirals at $z=0.83$, and 25 field spirals at lower
redshifts and  we  plan to increase this sample by one order of magnitude. We
present here the first results of our study, and discuss the implications of
our data in the context of current ideas and models of galaxy formation and
evolution.

\end{abstract}

\end{opening}           

\section{Introduction}

Ground-based and HST observations indicate that the disk galaxy population in
rich galaxy clusters has experienced remarkable evolution since $z = 1$. It has
been argued that the increase with time of the S0 fraction and the simultaneous
decrease in the spiral fraction suggest star-forming spirals fall into distant
clusters at much higher rates than nearby, and that these spirals ultimately
become S0s when star formation is extinguished by the cluster environment. 
Recent hydro-dynamical simulations of the interaction of the gaseous components
of disk galaxies with the intracluster medium support these ideas (Quilis,
Moore \&  Bower 2000). 

The strong evolution of the cluster spiral population contrasts with
the mild evolution observed in the field spirals to $z\sim 1$ (cf.\ Vogt
2000, and references therein). To quantify the evolution of the cluster
spirals, we are measuring the stellar and dynamical masses and $M/L$ ratios
for a sizeable sample of morphologically-classified disk galaxies in rich
galaxy clusters at $0.2 \! < \! z \! < \! 0.9$.  We present here the first
results.

\section{Rotation curves and the Tully--Fisher relation}

The first cluster we have studied is MS1054$-$03 at $z=0.83$  (van Dokkum et
al.\ 1999).  Multi-object spectroscopy was obtained using  FORS2 at the
VLT\@.     We selected 11 cluster members which were known to have   [OII]3727 
emission and 2 cluster members with  disk morphologies but without known
emission characteristics. For comparison purposes,  the rest of the slits were
placed on galaxies with spiral morphologies, and with magnitudes and colours
similar to the known cluster disk galaxies. Since we wanted to measure rotation
curves, the slits were aligned with the major axes of the galaxies.  Two
multi-slit masks were made  at right angles to maximise the range of 
major-axis position angles that could be covered.  The total exposure times
were 3.5~hours per mask.  We found evidence for rotation (tilted emission
lines)  in 9 of the known $z=0.83$ cluster galaxies, plus in 1 newly found
cluster galaxy. In addition rotation was seen in about 25 foreground galaxies
and 1 background galaxy. These field spirals provide a random field galaxy
sample, observed with the same instrument and under the same observing 
conditions, and thus ideally suited for direct comparison. 

Rotation curves were derived for the brighter emission lines in the sample
from Gaussian fits at each spatial point along the slit (Figure~1). For all
the emission lines, $V_{\rm rot} \sin i$ was also estimated visually from the
2-D spectra. The rotation curves in Figure~1 have  not been corrected for the
effect of seeing and slit-width. We are in the process of employing the
synthetic rotation curve technique of Simard \& Pritchet (1999) to extract
$V_{\rm rot} \sin i$ from the 2-D spectra taking into account these effects.

\begin{figure}
\centerline{\includegraphics[width=15cm]{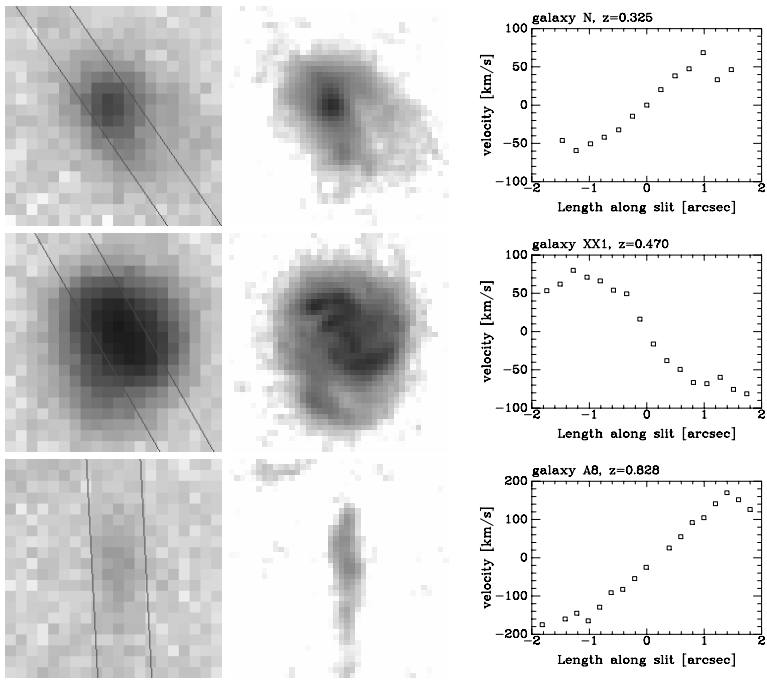}}
\caption{Three examples of the galaxies for which we have
obtained rotation curves. 
From left to right the panels show:
FORS2 R-band images ($4'' \times 4''$ section, 0.6$''$ seeing)
with the 1$''$ slits overlayed;  
the WFPC2 F606W images; and 
the preliminary VLT rotation curves 
with no correction for the line-of-sight inclination of the galaxy,
nor for the effect of the slit-width and seeing.
The first two galaxies are field galaxies, and the third one is one of our 
faintest cluster galaxies. Their 
magnitudes are $R$ = 21.9, 20.8 and 24.0 from top to bottom. 
}
\end{figure}

Figure~2 shows a preliminary Tully--Fisher relation for a subsample of
our galaxies. To ensure homogeneity,  we only include 
morphologically-classified spiral galaxies (from HST imaging)  for which the
[OII]3727 line was observed  (this excludes about 15 low $z$ foreground
galaxies).  This subsample contains 7 foreground galaxies ($z =
0.43$--0.76), 8 cluster galaxies ($z = 0.83$), and one background
galaxy ($z = 0.90$).

\begin{figure}
\centerline{\includegraphics[width=18cm]{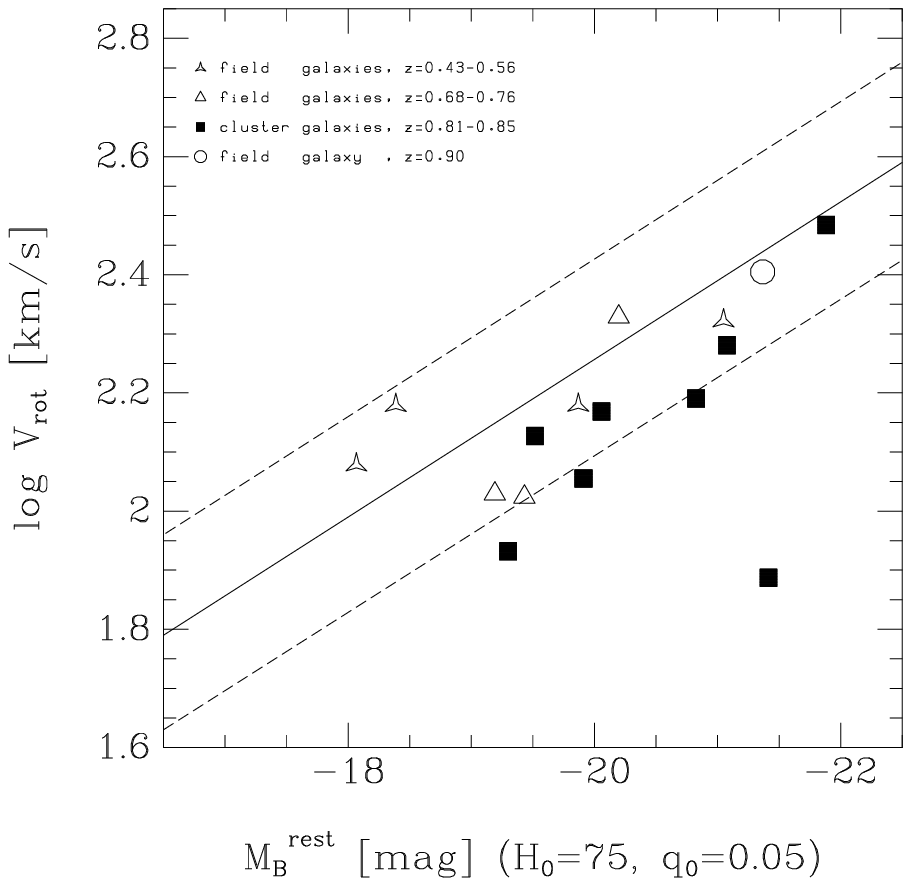}}
\caption{
Preliminary high redshift cluster and field Tully--Fisher relation. The solid
line shows a local HI based TF relation for 32 cluster spirals from Pierce and
Tully (1992), cf.\ Vogt (2000). The dashed lines show the 3 sigma limits.  
Filled symbols are cluster spirals, and open symbols field ones. The deviating
cluster galaxy has uncertain morphological classification: edge-on disk with
dust at the centre, or two almost overlapping edge-on disks.
}
\end{figure}

Total magnitudes and ellipticities were measured  in HST+WFPC2 F814W images
(van Dokkum et al.\ 1999).  The F814W magnitudes were transformed to 
the rest-frame $B$-band (Fukugita et al.\ 1995), and 
absolute magnitudes were calculated using $H_0 = 75\,{\rm km}\,{\rm
s}^{-1}\,{\rm Mpc}^{-1}$ and $q_0 = 0.05$. No correction for internal
extinction was applied. $V_{\rm rot} \sin i$ values were  derived from the
observed resolved [OII] emission lines, and  $\sin i$ was calculated from the
ellipticities measured in the HST images.

\section{Discussion}

Figure~2 shows that at a fixed $V_{\rm rot}$ the high $z$ cluster galaxies
appear to be brighter on average than the high $z$ field galaxies, and than the
local relation. It is tempting to interpret this as the result of enhanced star
formation on spiral galaxies falling onto the cluster. Our sample
preferentially contains star forming spirals since we selected  emission line
galaxies from a rest frame $B$-magnitude limited sample. One could speculate
that after this initial episode of enhanced star formation, the interaction
with the intergalactic medium will remove much of the gas in these galaxies
(cf.\ Quilis et al.\ 2000), and the star formation will cease. After an E+A
phase the spirals could turn into S0s. However, we must stress that  a larger
cluster sample, covering a range of redshifts,  a more careful analysis of the
data and detailed modelling are necessary to reach firm conclusions.

\acknowledgements


Generous financial support from
the Royal Society (AA-S) and the Danish Research Training Council (BM-J)
is acknowledged.

\end{article}
\end{document}